\documentclass[11pt,twoside]{article}

\usepackage{asp2006}
\usepackage{epsf}
\usepackage{lscape}

\begin{document}
\title{Variations of the Radio Synchrotron Spectral Index in M33}   %%% Fill in title
\author{F.~S. Tabatabaei\altaffilmark{1}, R. Beck\altaffilmark{1}, E. Kr\"ugel\altaffilmark{1}, M. Krause\altaffilmark{1}, E.~M.~Berkhuijsen\altaffilmark{1}, K.~D. Gordon\altaffilmark{2}, K.~M. Menten\altaffilmark{1}}   %%% Fill in author names
%\affil{}    %%% Fill in author affiliations
\altaffiltext{1}{Max-Planck Institut f\"ur Radioastronomie, Auf dem H\"ugel 69, 53121 Bonn, Germany}
\altaffiltext{2}{Space Telescope Science Institute, Baltimore, MD 21218}
\begin{abstract} %%% Abstract to run on from here.
We   determine the  variation  in  the nonthermal   radio  spectral  index  in  the  nearby  spiral galaxy   M33.   We separate the thermal and nonthermal components of the radio continuum emission without the assumption  of  a  constant   nonthermal  spectral  index. Using  the  Spitzer  FIR  data  at  70  and  160\,$\mu$m  and a standard  dust  model,  we  de-redden  the  H$\alpha$ emission.  The  extinction-corrected   H$\alpha$  emission   serves   as   a template   for   the   thermal   free-free   radio   emission.  Subtracting from  the observed  3.6 and 20cm  emission (Effelsberg  and  the  VLA)   this  free-free  emission,  we obtain  maps  of  the  nonthermal  intensity  and  spectral index.
\end{abstract}

%%% MAIN BODY OF TEXT GOES HERE. CONSULT "INSTRUCTIONS FOR AUTHORS USING
%%% LATEX2E MARKUP", SECTIONS 2.3-2.6 FOR HELP WITH EQUATIONS, FIGURES,
%%% AND TABLES.
\section{Introduction}
The problem of separating the two components of the radio continuum emission, free-free (thermal) and synchrotron (nonthermal) emission,  dates back to the beginnings of radio astronomy. The usually applied technique is based on  the assumption of a constant nonthermal spectral index.  Although this assumption may be reasonable for global studies, it does not lead to a correct thermal/nonthermal distribution in detailed studies, and it is not possible to investigate the origin and energy-loss processes of the electron component of cosmic rays (CRs).
%We obtained the distribution of the dust optical depth at the H$\alpha$ wavelength  for the whole M33 by analyzing dust emission and absorption using the high sensitivity and resolution Spitzer (MIPS) FIR data at 70 and 160\,$\mu$m. This led to an H$\alpha$ map corrected for extinction, our free-free template.

\section{Separation of Thermal and Nonthermal Emission from M33 }
%{\bf Dust temperature and optical depth.}
The  dust color   temperature obtained between  MIPS 70\,$\mu$m  and  160\,$\mu$m (Tabatabaei et al. 2007a)  shows  variations between    19\,K  and  28\,K with a most  probable value  at 21.5\,K. Using the 160\,$\mu$m column density and the dust temperature maps, we mapped the dust optical depth at 160$\mu$m and then at H$\alpha$ wavelength ($\tau_{{\rm H}{\alpha} }$). 
In the extended central region and in the two main arms IN and IS, $\tau_{{\rm H}{\alpha} }>$\, 0.5 and between 0.2 and 0.4 in other arms.  
We found a radial gradient for the mean extinction in rings given by: $\tau_{{\rm H} _{\alpha}} ({\rm R})= (-0.009 \pm 0.002)\,{\rm R} + (0.24 \pm 0.03)$ .
%{\bf Thermal and nonthermal emission.}
Assuming a homogeneous distribution of the emitting and attenuating medium in the line of sight (for $\sim$\,0.4\,kpc spatial resolution), we de-reddened the observed H$\alpha$ emission (Tabatabaei et al. 2007a).
We found that less than 25\% of the total H$\alpha$ emission is obscured 
by dust  in M33.  The  intrinsic  H$\alpha$  emission measure converted to the thermal radio emission (assuming recombination case-B, e.g. Dickinson et al. 2003). Subtracting the thermal emission from  the observed  3.6\,cm and 20\,cm emission (Tabatabaei et al., 2007b), the nonthermal  maps were derived. We   obtained a  total  thermal fraction   of $\simeq$\,17\% at  20cm and  $\simeq$\,50\%  at 3.6\,cm  for galactocentric radii R$<$\,7.5\,kpc. The nonthermal emission   is more  smoothly  distributed   than the   thermal  emission. Furthermore, the structure of the  nonthermal  emission is smoother at 20\,cm than at  3.6\,cm,  indicating  a  stronger diffuse component of the synchrotron  emission  (in the  form of a  diffuse disk or halo)  at 20\,cm (see also Tabatabaei et al. 2007c).

\section{Distribution of the Nonthermal Spectral Index}
%{\bf Nonthermal spectral index.}
Using  the  nonthermal  maps at 20 and 3.6\,cm,  the  nonthermal  spectral  index map  was  obtained.  In the star forming regions,  the   nonthermal   spectrum  is relatively flat with an  average  value  of  0.6\,$\pm$\,0.1,  the  typical  spectral  index of supernova remnants, but  it increases to 1.2\,$\pm$\,0.2  in   the  interarm  regions  and outer  parts  of  the  galaxy.  This indicates energy losses of the relativistic electrons while they diffuse away from their origin in starforming regions towards the interarm regions and the outer parts of the galaxy. For the first time, a nonthermal spectral index map can be used to achieve more realistic models for the propagation of CR electrons. Figure.~1 shows that the  total   radio emission  is  mostly  nonthermal  at  R\,$>$\,4.5 kpc in M33,  where  the  spectral  index  is  dominated  by  synchrotron and  inverse Compton loss.

\begin{figure}
\begin{center}
\resizebox{6cm}{!}{\includegraphics*{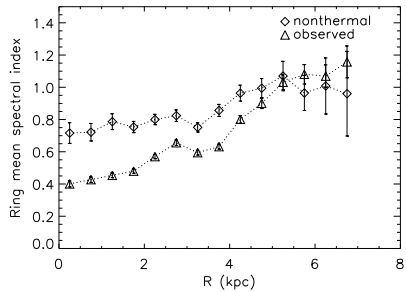}}
\caption[]{ Mean nonthermal spectral index in rings of 0.5\,kpc width in the galactic plane versus galactocentric radius. The total (observed) spectral index is also shown for comparison. The minima at $3<$R$<4$\,kpc is due to the HII complex NGC604.  }
\end{center}
\end{figure}

%\section*{}    %%% Unnumbered top level section head (remove "%" symbol)
%\subsection*{}   %%% Unnumbered second level section head (remove "%" symbol)

%\acknowledgements %%% Text of acknowledgements runs on after this command.

%%% THE BIBLIOGRAPHY
%%%
%%% CONSULT SECTION 3 OF "INSTRUCTIONS FOR AUTHORS" FOR HOW TO USE NATBIB.
%%% AUTHORS ARE ENCOURAGED TO USE EITHER THE "THEBIBLIOGRAPY" ENVIRONMENT
%%% BY UNCOMMENTING (DELETING THE "%" SYMBOL) THE COMMANDS BELOW, OR BY
%%% USING THE BIBTEX ENVIRONMENT. TO FIND OUT WHICH IS APPLICABLE TO YOUR
%%% CONTRIBUTION, CONSULT THE VOLUME EDITORS FOR YOUR PROCEEDINGS.
%%%

%\bibliography{s.bib}
\end{document}